\documentclass[twocolumn,preprintnumbers,aps,prb,final,amsfonts,amssymb,amsmath,floatfix,showpacs]{revtex4}

\usepackage[]{graphicx}
\usepackage{epstopdf}
\usepackage{bm}

\newcommand{\eg}{e.g.,}

\begin{document}
\title{On conversion of luminescence into absorption and the van Roosbroeck-Shockley relation}
\author{Rupak Bhattacharya}
\affiliation{Indian Institute of Science Education \& Research--Kolkata, Mohanpur Campus, Nadia 741252, West Bengal, India}
\author{Bipul Pal}
\affiliation{Indian Institute of Science Education \& Research--Kolkata, Mohanpur Campus, Nadia 741252, West Bengal, India}
\author{Bhavtosh Bansal}\email{bhavtosh@iiserkol.ac.in}
\affiliation{Indian Institute of Science Education \& Research--Kolkata, Mohanpur Campus, Nadia 741252, West Bengal, India}
\preprint{Applied Physics Letters (2012)}
\date{\today}

\begin{abstract}
The problem of conversion of experimentally measured luminescence spectrum into the absorption cross section is revisited. The common practice of using the van Roosbroeck-Shockley (or Kubo-Martin-Schwinger or Kennard-Stepanov) relation in this context is incorrect because luminescence from semiconductors is essentially all due to the spontaneous emission component of the recombination of carriers distributed far-from-equilibrium. A simple, physically consistent, and practical prescription for converting the luminescence spectra into absorption is presented and its relation to the so-called nonequilibrium generalization of the van Roosbroeck-Shockley relationship is discussed.
\end{abstract}

\maketitle

Quantum mechanics of interaction of the band states of a semiconductor with the photon field is well established and one has a choice of theories at different levels of sophistication to model the emission and the absorption process.~\cite{van roosbroeck, lasher-stern, pankove, bebb-williams, Smith, Basu, Seeger, Chatterjee, schubert, haug} If the non-radiative processes and various mechanisms of Stokes' shift\cite{StokesShift} can be neglected, then at least in principle, it should be straight-forward to infer the absorption coefficient of the semiconductor from the emission spectrum using the Einstein $A$ and $B$ coefficients~\cite{Seeger} or the microscopic theory.~\cite{lasher-stern, haug}

A robust scheme to convert the photoluminescence (PL) spectra into absorption is attractive. While absorption measurements with high dynamic range are difficult for both quantum wells as well as thick bulk samples, the emission (PL, electroluminescence) spectrum is much easier to measure over a large dynamic range. Indeed PL-excitation (PLE) spectroscopy has been a standard technique to infer the absorption coefficient for quantum wells. It will be nice if PL, which is a simpler measurement than PLE, can itself be used to infer the absorption. Also, it is the absorption cross section that is the theoretically calculated material property and one may want to compare it with the measured PL.~\cite{Chatterjee}

Long ago, assuming that in equilibrium with background thermal radiation, the photon emission rate $R(\hbar \omega)$ from a body would simply be the rate of absorption $\alpha(\hbar \omega)$ of thermal photons, van Roosbroeck-Shockley (vR-S) derived the following relationship:~\cite{van roosbroeck, fnote}
%%%%%%%%%%%%%%%%%%%%%%%%%%%%%%%%%%%
\begin{equation}\label{v R-S}
R(\hbar \omega) = {{[\hbar \omega]^2 n^2} \over {\pi^2 c^2 \hbar^3}} \alpha(\hbar \omega) {1 \over {\exp(\hbar \omega / k_{B} T) - 1}}.
\end{equation}
%%%%%%%%%%%%%%%%%%%%%%%%%%%%%%%%%%%%%
This expression has become a standard prescription to relate the PL with the absorption spectra.~\cite{van roosbroeck, lasher-stern, pankove, bebb-williams, Smith, Basu, Seeger, Chatterjee, schubert, haug, kost, ultralow, Schenk, yu-cardona, Rosencher, Reinhart, ihara, Subashiev} The same prescription has also been used under the names of the Kubo-Martin-Schwinger (KMS)~\cite{Chatterjee, ihara} and Kennard-Stepanov relations.~\cite{ihara,fnote1} Clearly the use of Eq.~\eqref{v R-S} to extract the absorption from PL has a problem; as $T \rightarrow 0$, the emission rate $R(\hbar \omega)$ goes to zero. But there must be spontaneous emission even at zero temperature. The zero-point modes of the electromagnetic field have no temperature dependence. In fact, PL measurements are usually done at low temperature to reduce the non-radiative recombination contribution.

%%%%%%%%%%%%%%%%%%%%%%%%%%%%%%%%%%%%%%%%
\begin{figure}[htb]
\includegraphics[clip,width=6.5cm]{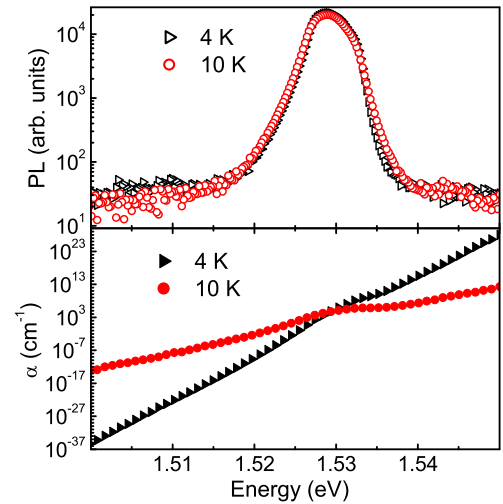}
\caption{(color online) Low-energy tails of the experimental PL (top) from GaAs quantum well at 4 and 10~K are almost same, whereas they seem to be drastically different when transformed into absorption (bottom) using vR-S relation}\label{fig1}
\end{figure}
%%%%%%%%%%%%%%%%%%%%%%%%%%%%%%%%%%%%%%%%%

Using Eq.~\eqref{v R-S}, values of the absorption coefficient have been inferred in a range that is orders of magnitude larger than the dynamic range of the corresponding PL spectrum.~\cite{ultralow} This point is clarified in Fig.~\ref{fig1}. There is no significant difference between the low-energy tails of the experimentally measured PL spectra at 4 and 10~K from a GaAs quantum well. But when these spectra are transformed into absorption (normalized to experimentally known value of the absorption coefficient at exciton energy~\cite{masumoto}) using Eq.~\eqref{v R-S}, a drastically different bandtail behavior is observed, with the extension of the bandtail states to ultralow values of the absorption coefficient. This artifact originates from the $\exp(\hbar \omega/k_{B}T)$-term in Eq.~\eqref{v R-S}. Through this term, temperature $T$ artificially controls the behavior (e.g., exponential tail and its slope) of the absorption spectrum derived from PL. Thus, one has to be careful in drawing any conclusion on Urbach edge %~\cite{urbach}
from PL data through the use of Eq.~\eqref{v R-S}.~\cite{kost}

Equation~\eqref{v R-S} is indeed a correct formula but it is often inappropriately handled. Actually, it is just a quantitative statement of Kirchhoff's law~\cite{Reif} of thermal radiation -- an imperfect absorber is also an imperfect emitter. A body in thermal equilibrium (within itself but not necessarily with the surroundings), like an incandescent lamp or a star with its own source of energy~\cite{Nair} will emit a black-body spectrum only if its density of states form a continuum and are energy independent. The energy-dependent `greybody' factor $\alpha(\hbar \omega)$ carries information about structure of the density of states.

It is erroneous to use this relationship to describe \emph{interband} emission from a semiconductor because this can not be described by thermal equilibrium (or even detailed balance) condition, regardless of how small the excitation intensity is. The emission peak (if it is $\sim 1$~eV) corresponds to a temperature of many thousand Kelvin and only the spontaneous emission process is important in such cases. Approximate thermal equilibrium only exists for intraband phenomena.~\cite{ultralow}

In this letter, we derive the relationship between emission and absorption spectra in a useful form using the standard framework of Einstein $A$ and $B$ coefficients without resorting to a detailed balance argument. While this treatment is very elementary, classic~\cite{pankove, Smith} and modern~\cite{schubert, yu-cardona, Basu, Rosencher} textbooks on semiconductor physics continue to derive the vR-S relation [Eq.~\eqref{v R-S}] and are at best ambiguous about its scope.

Consider optical transitions between two energy levels, $c$ and $v$ referring to states in the conduction and valence band, respectively. According to Einstein's theory of spontaneous and stimulated emission,~\cite{Seeger} the rate of downward transitions from the excited levels $c$ to the lower level $v$ is
%%%%%%%%%%%%%%%%%%%%%%%%%%%%%
\begin{gather}
R(\hbar \omega) = A_{cv} + B_{cv} D(\hbar \omega), \\
\text{where} \;\; D(\hbar \omega) = {{8 \pi n^{3} [\hbar \omega]^{2}} \over {h^{3}c^{3}}}{1 \over {\exp(\hbar \omega/k_{B}T)-1}}, \\
A_{cv} = {{8 \pi n^{3} [\hbar \omega]^{2}} \over {h^{3} c^{3}}}B_{vc},\;\; \text{and} \;\;B_{cv}=B_{vc}.
\end{gather}
%%%%%%%%%%%%%%%%%%%%%%%%%%%%%%%%%
Here, $D(\hbar \omega)$ is the photon distribution function and $B_{vc}$ is the rate of photon absorption in the medium.

Let us denote the absorption cross section, $\sigma(\hbar \omega)$, as the probability of absorption under the condition that the lower (higher) energy state is guaranteed to be occupied (unoccupied). By introducing a mean lifetime $\tau(\omega)$ of photons passing through an absorbing medium, $B_{vc}$ can be related to $\sigma(\hbar\omega)$.~\cite{pankove, Smith, Basu} With $B_{vc}=1/\tau(\omega)$ and $\tau(\omega)=n[c\sigma(\hbar\omega)]^{-1}$, we get $B_{vc}=\sigma(\hbar\omega)c/n$. Then putting everything together, we get a relationship between emission and absorption \emph{rates}:
%%%%%%%%%%%%%%%%%%%%%%%%%%%%%%%%%
\begin{equation}
R(\hbar \omega)={{n^2 [\hbar\omega]^2} \over {\pi^2 \hbar^3c^2}}\sigma(\hbar\omega)\left[1+{1\over \exp[\hbar\omega/k_{B}T]-1}\right].
\end{equation}
%%%%%%%%%%%%%%%%%%%%%%%%%%%%%%%%%
For most semiconductors, i.e. those with an energy gap of more than, say, 200 meV,   $\hbar\omega> 8 k_BT\Rightarrow 1+[\exp(\hbar\omega/k_BT)-1]^{-1} \sim 1$, even at room temperature. Hence it is only the first term in the square bracket which contributes to  emission and we end up with an expression, $R(\hbar\omega)\sim [\hbar\omega]^2 \sigma(\hbar\omega)$, drastically different from Eq. (\ref{v R-S}). Indeed this expression for spontaneous emission rate is the same as Eq.~19(b) in the classic paper by McCumber.~\cite{McCumber}

The actual photoluminescence spectrum is given by the product of the emission rate and the probability of the excited state being occupied and the ground state being unoccupied. Thus, the emission spectrum $\Im(\hbar \omega)$ and $\sigma(\hbar\omega)$ are related through:
%%%%%%%%%%%%%%%%%%%%%%%%%%%%
\begin{equation}\label{finalexpression}
\Im(\hbar\omega)={n^2 [\hbar\omega]^2\over \pi^2 \hbar^3c^2}\sigma(\hbar\omega)f_c(1-f_v).
\end{equation}
%%%%%%%%%%%%%%%%%%%%%%%%%%%%%%
Here $f_c=[\exp(E_c-E_F^c)/k_BT+1]^{-1}$ and $(1-f_v)=[\exp(E_F^v-E_v)/k_BT+1]^{-1}$ are the probabilities of the conduction band states being occupied and the valence band states being empty, $E_F^c$ ($E_F^v$) denotes the quasi-Fermi energy for electron (hole) distribution in the conduction (valance) band,~\cite{Anselm} and $E_c$ and $E_F^c$ are measured from the bottom of the conduction band whereas $E_v$ and $E_F^v$ are measured from the top of the valence band. The actual transition energy is $\hbar \omega=E_c-E_v$. An approximation to Boltzmann like form:
%%%%%%%%%%%%%%%%%%%%%%%%%%
\begin{equation}
f_c(1-f_v)=\exp\left[-{{\hbar\omega-E_g} \over {k_BT}}\right]\exp\left[{{\Delta F-E_g}\over{k_BT}}\right]
\end{equation}
%%%%%%%%%%%%%%%%%%%%%%%
with $\Delta F=F_c-F_v$ is only possible in the regions where $E_c-E^c_F>>k_BT$ and $E_F^v-E_v>>k_BT$ and not everywhere. Close to the band edge, the complete Fermi distributions should be used. This important point is often overlooked.~\cite{ultralow,kost}

\begin{figure}[htb]
\includegraphics[clip,width=5.0cm]{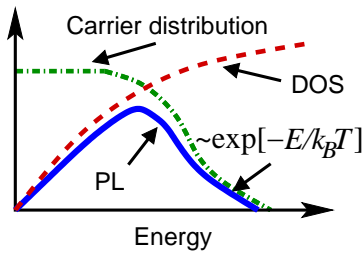}
\caption{(color online) Schematic of the physical content of Eq.~\eqref{finalexpression}. The PL signal is just the product of the absorption spectrum that is proportional to the joint density of states~\cite{pankove} and the {\it quasi}-Fermi distributions for electrons and holes. The high energy tail of PL is proportional to $\exp[-E/k_{B}T]$. For regions close to the bandedge, the complete Fermi distributions must be used.}\label{fig2}
\end{figure}

The above expression [Eq.~\eqref{finalexpression}] for the emission spectrum also matches with our intuition that a PL spectrum should just be the density of states at the low energy side and the high energy side should denote the distribution function (Fig.~\ref{fig2}). For example, it is a standard practice to fit the electron (Boltzmann) distribution to the high energy tail to extract the carrier temperature. Note that the carrier temperature parameterizes the distribution function within a single band for which the notion of equilibrium is approximately valid.

%%%%%%%%%%%%%%%%%%%%%%%%%%%%%%%%%%
%%%%%%%%%%%%%%%%%%%%%%%%%%%%%%%%%

One also encounters a `nonequilibrium van Roosbroeck-Shockley' relation:~\cite{lasher-stern, Basu, Epstein-Sheik-baha}
%%%%%%%%%%%%%%%%%%%%%%%%%%%
\begin{equation}\label{KMS}
\Im(\hbar \omega)={[\hbar \omega]^2n^2\over \pi^2c^2\hbar^3}\alpha(\hbar\omega){f_c(1-f_v)\over f_v-f_c}.
\end{equation}
%%%%%%%%%%%%%%%%%%%%%%%%%%%
It is also recognized that the vR-S relation can be derived from the above relation under equilibrium conditions and indeed $[f_c(1-f_v)]/(f_v-f_c)=[\exp(\hbar\omega/k_bT)-1]^{-1}$ when $\Delta F=0$ (at equilibrium).~\cite{Basu} Comparing Eq.~\eqref{KMS} with our Eq.~\eqref{finalexpression}, the difference is in the definition of the absorption coefficients in the two expressions: $\sigma(\hbar\omega)$ in Eq.~\eqref{finalexpression} is the absorption cross section, viz. the absorption coefficient under the condition that the conduction band is empty and the valence band is full, whereas $\alpha(\hbar\omega)$ in Eq.~\eqref{KMS} is the absorption coefficient under the same condition of nonequilibrium carrier density which existed for during luminescence\cite{Roucka}, $\sigma=\alpha/(f_v-f_c)$. Put in another way, in Eq.~\eqref{KMS} the absorption coefficient $\alpha$ is the rate of photon absorption under the condition of detailed balance between the emission and absorption process under a certain nonequilibrium electron and hole density (and thats why it yields the original vR-S relation for equilibrium). While Eq.~\eqref{KMS} is correct and indeed very useful in situations where one wants to model the physics when both absorption and emission processes are \emph{simultaneously} operational (for example, in determining the semiconductor laser gain threshold~\cite{Seeger}), this absorption coefficient ($\alpha$) is a strongly optical excitation (or electrical injection) dependent quantity and not a fundamental material parameter which one would normally be interested in, and what the authors of,  for example, Refs. ~\onlinecite{Chatterjee}, ~\onlinecite{kost}, ~\onlinecite{ultralow}, ~\onlinecite{ihara}, ~\onlinecite{Subashiev} had in mind. Theoretically also $\sigma (\hbar\omega)$ is the quantity that is usually calculated and not the nonequilibrium carrier density-dependent $\alpha(\hbar\omega)$ of Eq.~\eqref{KMS}.

To summarize, we have emphasized that usual van Roosbroeck-Shockley relation [Eq.~\eqref{v R-S}], though correct, is often inappropriately used in relating the absorption to the spontaneous emission spectra of a semiconductor. We demonstrated that another elementary expression [Eq.~\eqref{finalexpression}] accomplishes the task and is physically consistent. It is hoped that with this clarification, more studies can consistently utilize the information contained in the photoluminescence spectra by connecting it to the absorption cross section, which is a material property.


\begin{thebibliography}{99}
\bibitem{van roosbroeck} W. van Roosbroeck and W. Shockley, Phys. Rev. \textbf{94}, 1558 (1954).
\bibitem{lasher-stern} G. Lasher and F. Stern, Phys. Rev. \textbf{33}, A553 (1964).
\bibitem{pankove} J. I. Pankove, \emph{Optical Processes in Semiconductors}, (Prentice-Hall, Englewood Cliffs, NJ, 1971).
\bibitem{bebb-williams} H. Barry Bebb and E. W. Williams, \emph{Photoluminescence I: Theory} in Semiconductors and Semimetals, Vol. 8, (Academic, New York, 1972).
\bibitem{Smith} R. A. Smith, \emph{Semiconductors}, (Cambridge Univ. Press, 1978).
\bibitem{Basu} P. K. Basu, \emph{Theory of Optical Processes in Semiconductors}, (Oxford, New York, 1997).
\bibitem{Seeger} K. Seeger, \emph{Semiconductor Physics: An Introduction}, (Springer, Berlin, 2004).
\bibitem{Chatterjee} S. Chatterjee, C. Ell, S. Mosor, G. Khitrova, H. M. Gibbs, W. Hoyer, M. Kira, S. W. Koch, J. P. Prineas, and H. Stolz, Phys. Rev. Lett. \textbf{92}, 067402 (2004).
\bibitem{schubert} E. F. Schubert, \emph{Light-Emitting Diodes}, (Cambridge Univ. Press, 2006).
\bibitem{haug} H. Haug, A. P. Jauho,\emph{Quantum Kinetics for Transport and Optics in Semiconductors}, (Springer, Berlin, 2008).
\bibitem{StokesShift} B. Bansal, A. Kadir, A. Bhattacharya, B. M. Arora, and R. Bhat, Appl. Phys. Lett. {\bf 89}, 032110 (2006).
\bibitem{fnote} Standard symbols, {\eg}  $\hbar$, $k_{B}$, used throughout the text have their respective usual meaning, unless mentioned otherwise.
\bibitem{kost} A. Kost, H. C. Lee, Y. Zou, P. D. Dapkus, and E. Garmire, Appl. Phys. Lett. \textbf{54}, 1356 (1989).
\bibitem{ultralow} E. Daub and P. W\"urfel, Phys. Rev. Lett. \textbf{74}, 1020 (1995); E. Daub and P. W\"urfel, J. Appl. Phys. \textbf{80}, 5325 (1996).
\bibitem{Schenk} H. P. D. Schenk, M. Leroux, and P. de Mierry, J. Appl. Phys. \textbf{88}, 1525 (2000).
\bibitem{yu-cardona} P. Y. Yu and M. Cardona, \textit{Fundamentals of Semiconductors: Physics and Materials Properties}, (Springer, Berlin, 2001)
\bibitem{Rosencher} E. Rosencher and B. Vinter, \textit{Optoelectronics}, (Cambridge Univ. Press, Cambridge, 2004).
\bibitem{Reinhart} F. K. Reinhart, J. Appl. Phys. \textbf{97}, 123534 (2005).
\bibitem{ihara} T. Ihara, S. Maruyama, M. Yoshita, H. Akiyama, L. N. Pfeiffer, and K. W. West, Phys. Rev. B \textbf{80}, 033307 (2009).
\bibitem{Subashiev} A. V. Subashiev, O. Semyonov, Z. Chen, and S. Luryi, Appl. Phys. Lett \textbf{97}, 181914 (2010).
\bibitem{fnote1}Within thermal field theory using the properties of the KMS states, Eq. (1) is a manifestation of the fluctuation-dissipation theorem.  For a field theoretic proof of Eq.~\eqref{v R-S}, see Ref. ~\onlinecite{Nair}, p405-409.
\bibitem{Nair} V. P. Nair, \textit{Quantum Field Theory: A Modern Perspective}, (Springer, Berlin, 2005).
\bibitem{masumoto} Y. Masumoto, M. Matsuura, S. Tarucha, and H. Okamoto,  Phys. Rev. B \textbf{32}, 4275 (1985).
\bibitem{Reif} F. Reif, \textit{Fundamentals of Statistical and Thermal Physics}, (McGraw-Hill, New York, 1965).
\bibitem{McCumber} D. E. McCumber, Phys. Rev. \textbf{136}, A954 (1964).
\bibitem{Anselm} A. Anselm, \textit{Introduction to Semiconductor Theory}, (Mir, Moscow and Prentice-Hall, Englewood Cliffs, N.J. 1981).
\bibitem{Epstein-Sheik-baha} R. Epstein and M. Sheik-Bahae, \textit{Optical Refrigeration: Science and Applications of Laser Cooling of Solids}, (VCH Publishers, 2009).
\bibitem{Roucka} R. Roucka, J. Mathews, R. T. Beeler, J. Tolle, J. Kouvetakis, and J. Men\'endez, Appl. Phys. Lett. {\bf 98}, 061109 (2011).
\end{thebibliography}
\end{document}